# Cross-disciplinary integration towards sustainable construction

*Dorian A. H. Hanaor*

Materialeyes, Materials Innovation Consulting

**Faculty of Science and Engineering, Southern Cross University, East Lismore, NSW, 2480, Australia**

dorian@materialeyes.net

## Abstract

**The built environment is the basis for human experience in the world, yet its current practices contribute significantly to global greenhouse gas emissions, resource depletion, and environmental degradation. This chapter explores the critical role of cross-disciplinary integration in achieving sustainable construction, emphasising the need to bridge the gaps between material science, architecture, engineering, environmental studies, and social sciences. By examining past and present examples and trends we explore how holistic, systems-oriented approaches can drive innovation and sustainability. We examine how the siloed nature of modern disciplines has hindered progress, leading to fragmented outcomes and missed opportunities for circularity and resource efficiency. We present the case for a paradigm shift towards cross-disciplinarity, where stakeholders across sectors and disciplines collaborate to create cohesive solutions that balance technical and economic performance, environmental stewardship, and social equity. This chapter addresses the economic and regulatory barriers that impede the adoption of sustainable materials and practices, advocating for structural changes in education, industry, and policy and highlighting the tools that are now enhancing this integration of knowledge. Ultimately, this work underscores the transformative potential of cross-disciplinary integration in redefining the built environment, offering a roadmap for achieving sustainability through collaboration, innovation, and a shared commitment to planetary well-being**.

### Introduction

During the Imperial period of Rome (27 B.C. – 4th Century A.D) the production of building materials underwent tremendous advancement [1]. Through the well informed extraction and combination of suitable raw materials and their physical and thermal processing, industrial and centralised production of bricks, glass and concrete allowed structural achievements in Rome itself, and throughout the empire, surpassing anything seen in comparable civilisations at the time [2]. Roman concrete is to this day an object of wonder and inspiration, enabling structures of unprecedented complexity and size. The processing – microstructure – performance relationships of roman concrete are to this day the subject of investigations, and findings have inspired a range of recent innovative ventures including among others DMAT [3] and Silica-X [4]. While perhaps lesser known, industrial roman brickmaking also allowed the expansion of efficiently constructed and well-built structures around the empire, using masonry units of consistent and standardised dimensions [5]. The centralised industrial production of materials for European built environments largely diminished with the decline of the Roman Empire.

Between the fall of the Roman Empire and the emergence of industrial societies in the early 19$^{th}$ century, vernacular construction prevailed, with construction of dwellings, places of worship, castles and fortifications being conducted as projects within communities. Typically, materials for construction were based on locally available resources. Wood, stone, lime-based cements and bricks were produced from regionally varying earth, rocks and biomass [6]. With the industrial revolution a paradigm shift occurred and the production of bricks, cement, timber elements, iron and later steel became the domain of centralised industrial production. Bricks in particular are typical of this change. In vernacular construction, hand-made loam based bricks were often used unfired, while fired clay bricks were produced locally in temporary clamp-kilns or simple intermittently fired bottle-shaped kilns [7]. Bricks produced in this manner are often inconsistent in their dimensions, and are not conducive to mass construction of structures. Unlike roman bricks and later industrial bricks that dominate the architectural landscape of much of Victorian London.

The shift to industrial production of materials in general, and building materials in particular, like many other trends of the 19$^{th}$ century, gave rise to increasingly narrowly defined disciplines within the realm of engineering and materials science and a diminishment of the role of local trades and craftspeople in shaping the material aspects of built environments. The producers of construction materials became increasingly disconnected from the broader contexts of the sourcing of raw materials and their implementation in built environments, which were often distant from the location of production. Standardised products have long since become the norm, and this has naturally provided advantages for architects and engineers seeking to design structures based on known quantities of material performance. However the disconnect between environmental custodians, materials producers and construction professionals has made it difficult to improve lifecycle outcomes and human experience in the production and use of materials.

A stark example of the emergence of new disciplines and the resulting disconnect from vernacular construction can be found in the field of cement chemistry. In the most typical production of cement clinker, limestone and clay rich minerals are milled and fired at very high temperatures in clinker kiln. The resulting carbon emissions, at least 60% of which arise due to the calcination of limestone, are a major obstacle towards sustainability of our built environments worldwide. Using a multitude of honed analytical tools such as the Bogue calculation, the Lea and Parker equations and numerous others, the cement chemist knows how to tailor raw materials and processing towards the achievement of desired microstructures, phase assemblages and cement performance [8-10]. However, the isolation of the discipline from geologists, environmentalists and environmentally-conscious architects and stonemasons means that the use of quarried limestone directly as masonry material is not considered by cement chemists and does not feature widely in the modern construction landscape, despite the improved sustainability outcomes offered by the use of quarried stone and the aesthetics and value experienced by inhabitants of stone dwellings [11-13].

Addressing climate change and biodiversity loss is emerging as a rallying point for human activity, as the awareness of the gravity of these challenges grows. Even with complete elimination of carbon emissions, it has become clear that carbon dioxide removal (CDR) is further essential to avoid catastrophic environmental change. To effectively align human activities with our environmental objectives, disciplines need to operate in synergy with one another to develop a systems thinking oriented approach to achieving human development goals in a sustainable manner [14].

Systems thinking enables a holistic understanding of the many interconnected factors influencing sustainability in the built environment. By bringing together the diverse stakeholders in this sector and viewing the built environment as a dynamic system, we can get a better view of how materials choices impact resource extraction, manufacturing, energy consumption, structural performance and lifecycle impacts [15,16].

**Disciplinary Silos**

The breadth of our economy means that much of the economic activity that takes place occurs in insulated silos. More specifically, producers of goods and suppliers of services are less aware in general of the broader context in which their economic activity takes place and are primarily occupied with optimising performance within in a limited field of focus. In the analysis of sustainable economies, the integration of disciplines and bridging of scales have been determined as key factors in delivering outcomes relating to industrial symbiosis, and in particular circularity, as well as enhancing the effectiveness of climate action and sustainable development [17-19]. Methodological and technological tools, including IoT, big data, game theory, network modelling and numerous others have been developed to expand system boundaries, bring together numerous input and output metrics from diverse stakeholders and provide disciplinary-integrating insights towards improved holistic industrial and environmental outcomes in many sectors of industry [19-21]. The study of industrial ecologies as complex systems exhibiting emergent properties underpins many lifecycle assessments and material flow analyses [22,23].

Expanding cross disciplinary integration in sustainable construction to extend beyond metrics of Greenhouse gas (GHG) emissions and create truly inclusive decision making processes, is rendered complex by the diversity of materials performance indicators, the complex environmental and social contexts in which the value chain of construction industry takes place and the consequent subjectivity in defining an effective functional unit in LCAs relating to materials and built environments [24-27]. The fuzzy nature of many metrics, and the difficulty in weighting human experience and social impacts has motivated the adoption of concepts of the built environment as social-ecological systems, [28-30]. These studies highlight the need to integrate cultural and natural elements across spatial scales from the national to the local scale to achieve improved social-LCA outcomes and identify the intricacies and assumptions that need to be considered when addressing a people-focused approach to sustainability and the built environment.

The more complex social intricacies of built environments notwithstanding, even within the confines of the industrial realm the optimisation of processing of materials towards use in the built environment is also carried out within silo-constrained closed loops, sensitive to performance and cost-effectiveness measured by discipline-specific metrics. The lack of transparency between supply chains in the construction industry and the absence of systems thinking are seen as barriers to disciplinary integration and efficient outcomes in the construction materials sector [15,31]. As with other industries, expanding the scope of symbiosis is achieved by developing holistic understandings of interconnected industries and identifying opportunities for circularity and process optimisation by integration across primary and secondary industries. For example, greater opportunities for industrial symbiosis and resulting circularity exist when processes in metallurgy and cement production are viewed holistically as opposed to separate loops of production and consumption. In particular tailings, metallurgical slags and overburdens from mining and metallurgy offer a rich menu of ingredients to produce low-carbon cements, Supplementary Cementing

Materials (SCMs) and alternative aggregates [32-36].

Taking a broader view further inclusivity can be achieved by integration with silos beyond primary and secondary industries, extending to the disciplines such as architecture and engineering that drive the use of materials in structures, and community planners who help shape the role of structures in built environments, and further afield to other stakeholders influencing the interface between humans and the built world. Some disciplinary silos are illustrated in Figure 1, showing the stakeholders and fields of focus within these silos. Naturally one can define silos with a higher degree of granularity and around the world a certain degree of integration is generally present between actors in these silos, which are naturally not totally isolated from one another.

Architecture in particular serves a prominent role in acting as a nexus point for disciplinary integration and indeed interdisciplinary design features in architectural education [37,38]. It is in the realm of architecture where the technical world is brought into harmony with human-experience at the structure-level. Waste management too sits at a strategically advantageous point, being a key enabler of the integration of industries and the built environment. For example, the use of waste products from agriculture as construction materials including concrete presents inspiring examples of industrial symbiosis that can be achieved when diverse silos are brought together. [39,40]. Also shown in Figure 1, Venture Capital, despite exhibiting its own finance-centric silo-mentality, also serves as a nexus point and acts as a conduit for economic catalysis of collaborative ventures, provided these produce a realisable financial advantage.

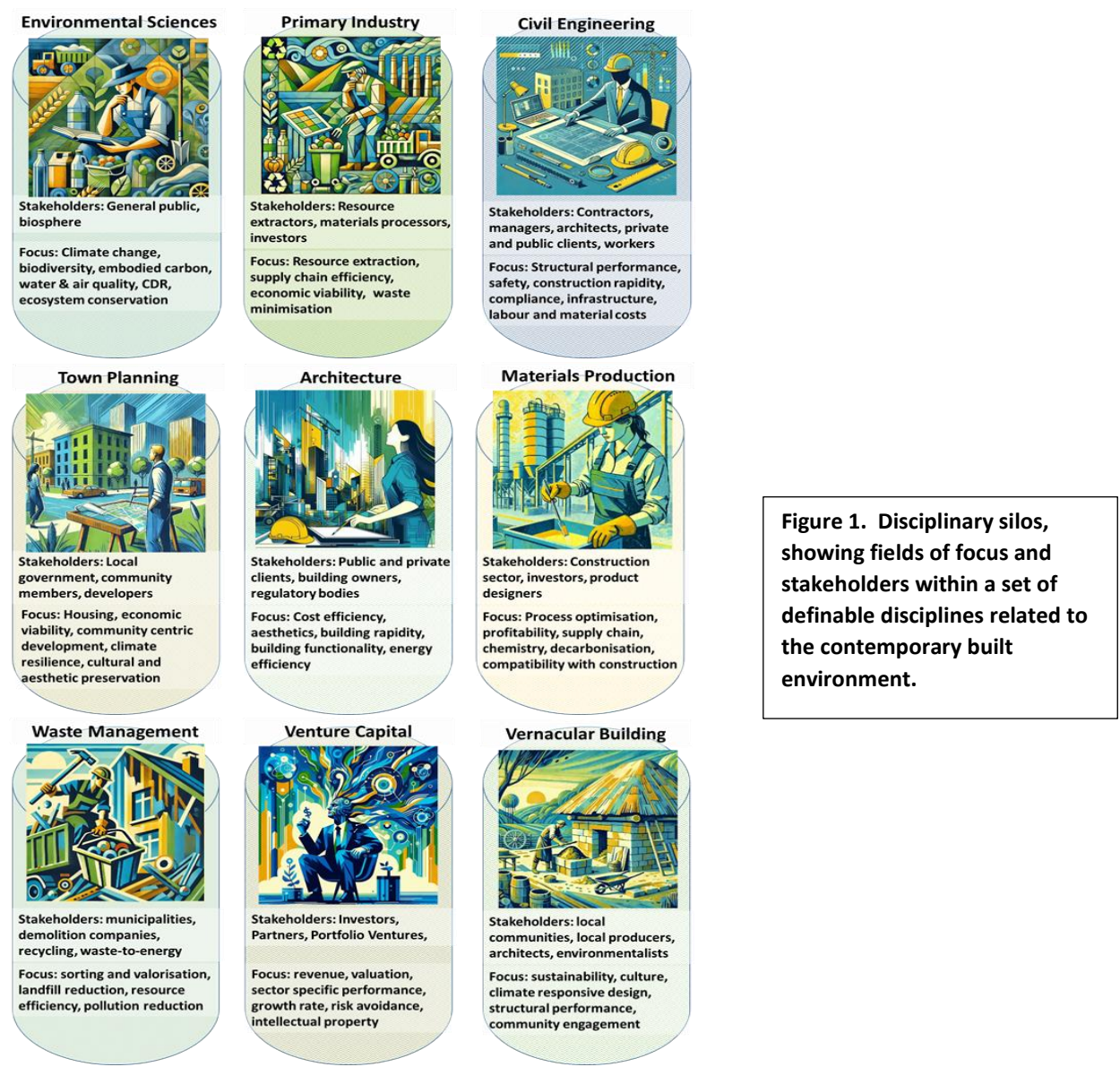

Figure 1. Disciplinary silos, showing fields of focus and stakeholders within a set of definable disciplines related to the contemporary built environment.

As mentioned, acknowledging the interconnectedness and complex-system nature of production and consumption of resources in the materials sector is a key step towards improving outcomes in the built environment and its impact on spheres of the environment, which themselves are highly interconnected. A systems thinking approach to building materials supply is accepted is being important towards robust and sustainable built environment value chains [41,42]. A systems thinking approach to building materials supply and use recognises that every stage—from extraction and production to transportation and disposal—is interconnected, and therefore decisions made at one point ripple through the entire value chain. This perspective, built upon the interconnection of the silos shown in Figure 1, allows a better understanding of the complex trade-offs involved in material selection, design, and construction processes.

Transcending the disciplinary silos and enabling systems thinking is also valuable towards addressing other 21$^{st}$ century challenges relating to the built environment in addition to sustainability. Thus for example heritage conservation, mass migration, conflict and disaster response may pose together conflated challenges towards the design of sustainable built environments that necessitate integrative systems thinking [43-46].

Moreover, the traditional separation of disciplines often fails to account for cultural, historical, and geographical contexts that influence material use. For example, vernacular construction practices, which often embodied deep local knowledge about material sourcing and environmental adaptation, have been largely sidelined in favour of standardised industrial materials. This disconnect limits opportunities to integrate traditional knowledge with modern advancements to achieve both sustainability and cultural relevance.

To break free from the confinement of these disciplinary silos and address the complex challenges in the construction material ecosystem, there is a pressing need for continued improvement in cross-disciplinary collaboration and systems thinking. As we have seen, the interdependencies between material production, design, policy, and environmental impact demand integrated solutions. Research has shown that siloed approaches often lead to fragmented outcomes, whereas cross-disciplinary collaboration frameworks and competencies in academia and industry foster effective innovation and sustainability [47,48]

By fostering dialogue and partnerships across the spectrum of stakeholders in the construction materials ecosystem—ranging from raw material suppliers and manufacturers to architects, urban planners, policymakers, and environmental advocates—we can create a shared vision for sustainable innovation. This requires not only structural changes within organisations and academic institutions but also the cultivation of a mindset that values interdisciplinary learning and co-creation. Educational programs, research initiatives, and industry practices must prioritise holistic thinking, encouraging professionals to understand the downstream and upstream implications of their decisions. This mentality is gaining increasing traction in educational domains, where programs that integrate engineering, environmental science, and social sciences have been shown to better prepare students for the complexities of sustainable development [49,50]. However, when it comes to the industrial and governmental implementation of integrated strategies for the production and usage of sustainable construction materials, significant barriers still remain.

**Cross disciplinary Integration**

The concept of bringing multiple disciplines together in the context of engineering challenges is well known and is a key feature of Systems Thinking and contemporary design philosophies. The need to bring multiple disciplines together in the identification and implementation of sustainable materials and practices is now well recognised towards various objectives, including materials circularity [51], heritage management [52], architectural design [53], energy efficiency [54] and many others. To a certain extent, nearly all projects and studies involve some integration of multiple disciplines and therefore it is important to highlight the uniqueness of the built environment in the demands it places on the breadth and depth of the fields that need to be brought under one roof. As the economic activities associated with construction and the usage of structures account for around 40% of global greenhouse gas emissions [55] and are a main driver of resource usage [56], the need to expand the integration of disciplines is particularly pressing in this sector.

In academic discourse, distinctions are often made between multi-disciplinarity, inter-disciplinarity, and trans-disciplinarity. Multi-disciplinary approaches involve parallel efforts by experts from different fields, while inter-disciplinary collaboration seeks to blend insights from these fields to address shared challenges. Trans-disciplinarity, however, goes a step further, creating a unified framework that transcends traditional disciplinary boundaries to foster co-creation and innovation. In the context of construction technologies, cross-disciplinary integration aligns closely with the principles of trans-disciplinarity, emphasising the synthesis of knowledge rather than its parallel application [57]. This distinction highlights the shift from merely involving multiple disciplines to truly integrating them into a cohesive, systems-oriented approach [58].

**Applied disciplinary integration in the built environment**

Cross-disciplinary integration has historically been a ubiquitous aspect of construction endeavours, evident even in early human societies. However, the degree to which this integration influences the nature and outcomes of construction activities has varied significantly over time. In today's construction landscape, characterised by siloed industry practices, fragmented planning processes, misaligned economic incentives, and the predominance of consumer-centric paradigms, the need for enhanced cross-disciplinary collaboration is increasingly apparent. Addressing these systemic challenges requires fostering integration across technical, economic, and operational domains to unlock efficiencies, drive innovation, and achieve sustainability objectives. Examples from our past, present and future can be found, highlighting the advantages that cross-disciplinary integration can bring to the realm of sustainable construction and inspiring the expansion of this integration.

Pullman, a region in Chicago built rapidly as a centrally planned industrial community, provides excellent examples for the integration across multiple disciplines towards industrial symbiosis and effective outcomes in the built environment [59-61]. Initiated by George Pullman in the 19$^{th}$ century, the construction of this community featured the integration of materials engineering, town planning, energy systems, residential construction and manufacturing to name but a few. The Clay used as the raw materials for the high quality bricks used to construct Pullman was sourced from the waste from the dredging of nearby lake Calumet, done to facilitate the transport of iron ore to the nearby steel mills, an early example of circular economy, while the factory operations were used to provide heat to the adjacent planned housing developments and waste water from residential communities was utilised as fertiliser in agricultural farms located near the brickyards. The efficiency implemented in the design, construction and operation of this community was made possible through cross-disciplinary integration, enabled by the centralised planning of this project.

Similar centrally planned projects including Masdar City in Abu Dhabi and Songdo Business District in Korea also are distinguished by an integration of fields to promote sustainable urban environments, with building sustainability metrics being certified in both cases by according to the LEED rating system (Leadership in Energy and Environmental Design), corresponding to aspects of lifecycle impacts in construction, operation and maintenance.

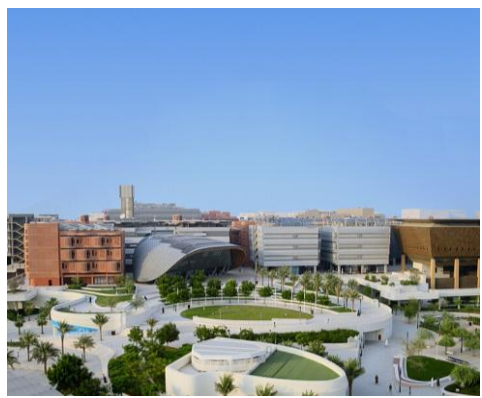
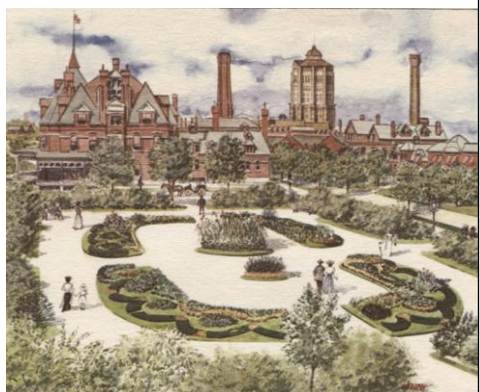

**Figure 2.** Centrally planned urban and urban/industrial developments. (left) photograph of Masdar City, in Abu Dhabi, UAE and (right) A watercolour painting of Pullman, Illinois, USA. Two examples of centralised planning across multiple disciplines enabling sustainable use of materials. Images available for use under the CC BY 4.0 license. [62,63]

Holistic approaches, often enabled by centralised planning, reduce barriers to systems thinking and foster disciplinary integration. Centralised planning allows for the alignment of diverse stakeholders, the harmonisation of goals, and the implementation of large-scale, innovative solutions that might otherwise be fragmented or constrained by competing interests. However, these barriers are harder to overcome in smaller-scale development projects, where the operational environment often prioritises cost-efficiency, risk mitigation, and adherence to established practices. This can lead to more conservative decision-making in the selection and implementation of construction materials and methods, limiting the potential for transformative innovation.

Despite these challenges, several pioneering projects and initiatives have demonstrated how holistic, systems-oriented approaches can be successfully applied across different scales and contexts. These examples illustrate the potential for integrating sustainability, innovation, and interdisciplinary collaboration into the built environment:

**BedZED** (Beddington Zero Energy Development), a residential development led by the Peabody trust and completed in 2002 in Sutton, south London integrates energy efficiency, renewable energy, and sustainable materials into a cohesive urban community that aims to offer all-round sustainability. By integrating professionals from a diverse range of fields the project stands out in enabling the use of local materials – over half materials were sourced within 50km of the project – and reclaimed materials – 15% of the total used [64,65]. The integrative multi-level planning of this development enable also sustainable lifecycle outcomes in terms of energy usage, housing affordability, water consumption, transport needs and community experiences [66,67]. The success of this development demonstrates the feasibility of reconciling creative and ambitious architecture with excellent social and environmental outcomes through cross-disciplinary integration. The total project cost including research and development and cost overruns, was £15 Million. As this project includes 100 homes and ~1000 m$^2$ of office space, the project demonstrates that sustainable outcomes are economically scalable [68].

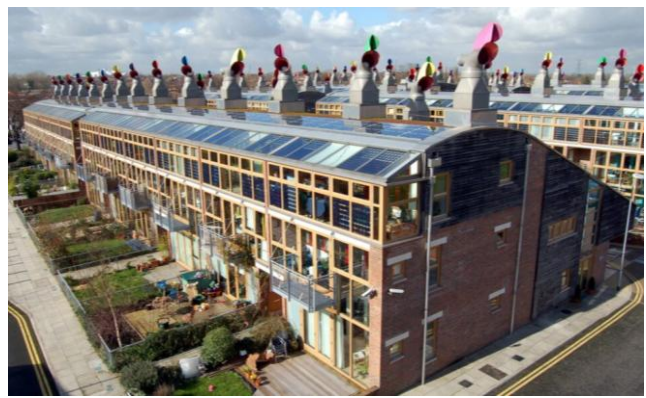

**Figure 3. BedZed (Beddington Zero Energy Development) Eco Village in the southern outskirts of London, England. Image available for use under the CC BY 4.0 license [69]**

**The Edge in Amsterdam:** designed by PLP Architecture and completed in 2014, the Edge is a commercial building that has been ranked by BREEAM as one the most efficient and sustainable commercial buildings worldwide [70]. In this development he integration of advanced smart materials, recycled materials, smart sensors and systems for energy generation and storage [71]. Through a sophisticated building management system (BMS) that draws on human feedback to optimise energy usage, and the installation of extensive renewable energy generation on the roof and façade, the building is able to generate more energy than it is consumed, with onsite storage further enhancing the utility of the energy generated. The demonstrated sustainability outcomes of this development show that by addressing challenges at multiple levels and integrating feedback from both humans and sensors, excellent outcomes can be achieved. This project is seen as a source of inspiration for office buildings of the future.

**The Ellen MacArthur Foundation** stands as a prominent example within a growing global network of organisations dedicated to accelerating the transition to a circular economy. Circularity networks such as this catalyse the transparency, collaboration, and systems thinking that are essential for achieving effective circularity across diverse industries, including the built environment. They act as conveners, knowledge hubs, and advocates, driving change at policy making, industry and consumer levels. In particular the Ellen MacArthur Foundation and other circularity networks work with cities and municipal waste handlers to implement circular economy strategies in the built environment, including promoting the reuse of building materials, modular construction and creating closed-loop systems for construction waste [72-74]. The New European Bauhaus (NEB) is a similar network that engages primarily on the community level and serves the role of bringing together stakeholders to reduce the barriers to industrial symbiosis and social, cultural and environmental sustainability outcomes.

Whether at small or large scales, it is clear that one of the main roles to be played by catalysts of cross-disciplinary integration in the built environment is lowering the barriers to communication and cooperation between disciplinary silos as well as addressing the contextual constraints imposed by the organisation of human activity – notably economic and regulatory.

**Economic and regulatory constraints**

The tendency for professionals across the various disciplinary silos illustrated in the preceding section to focus on their particular set of stakeholders, clients, and challenges stems largely from the structural and systemic pressures inherent in the economic and regulatory landscape. These constraints often incentivise short-term thinking, discipline-specific optimisation, and resistance to innovation—all of which can act as significant barriers to achieving holistic sustainability goals in the built environment.

*Economic Pressures and Market Dynamics*

In many cases, the economic incentives driving decisions in the built environment prioritise cost minimization and return on investment over long-term sustainability. Developers, for instance, often operate within tight budgets and timelines, which may lead to the selection of less expensive but environmentally detrimental materials. Similarly, materials producers may prioritise volume and cost efficiency over developing and scaling innovative, sustainable alternatives, given the high upfront costs and uncertain demand for such products. These economic pressures create a fragmented value chain where each stakeholder seeks to optimise for their immediate economic priorities rather than broader systemic goals.

The dominance of cost-driven decision-making presents a significant economic constraint to the adoption of sustainable materials in construction. This isn't simply a matter of clients aiming to save money; it's a complex interplay of market dynamics, risk aversion, and entrenched practices. Developers, facing pressure to maximise returns, may seek to minimise upfront costs, to maintain competitiveness in the market. This short-term focus may result in the selection of widely available materials with poor sustainability outcomes, and tends to overshadow the long-term benefits of sustainable materials, for which supply chains may not be as well-established. The widespread use of steel and OPC based concrete is particularly exemplary of this; despite the multitude of alternative materials with better LCA outcomes, including Cross Laminated Timber, Bamboo [75], Hempcrete [76] and Geopolymer concrete [77], steel and OPC based concrete prevail even in low-rise construction in many construction markets around the world. This is partly a result of the highly developed understandings of the processing and application of these materials. Metallurgy and cement chemistry have been intensively studied and developed through the lens of science and engineering for well over a century now. Within the confines of our industrial supply chain, research has shown that even minor adjustments to the selection process for materials, and giving further consideration to the application specific demands of the materials, can result in significant reduction in embodied carbon, without significantly higher costs [78,79]. Many of the poor practices in selecting and applying materials in the built environment can be attributed to an entanglement of education, labour experience and material availability (lead times). A feedback loop emerges in the market dynamics of construction sectors, whereby siloed practices reinforce one another, and create inertia in the technological landscape, which becomes resistant to the adoption of sustainable materials and practices.

This cost-centric approach is deeply ingrained in the fragmented structure of the construction industry. Multiple stakeholders—architects, engineers, contractors, subcontractors—operate under separate contracts with often conflicting incentives. The aforementioned siloed approach hinders collaboration and innovation. For instance, a contractor might be incentivised to minimise material costs within their specific scope, even if a slightly more expensive, sustainable material would lead to significant long-term savings for the building owner. The lack of integrated project delivery models and lifecycle cost analysis further exacerbates this issue. There's often no mechanism to capture and reward the long-term benefits of sustainable materials within the existing contractual frameworks, resulting in a so-called "Green Premium".

Supply chains for sustainable materials, while growing, often lack the aforementioned scale and maturity of those for conventional materials like concrete and steel. As long as sustainable alternatives remain sidelined it can be difficult to overcome barriers of longer lead times, and concerns about consistent availability. For example, while the demand for engineered wood products (EWPs) and the technology and understandings required for producing highly functional structural components from materials like Cross-laminated Timber (CLT) are developing, the relatively limited number of production facilities can create bottlenecks and price volatility [80-82]. This market immaturity creates a similar feedback loop in market dynamics, similar to the proverbial "chicken and egg" problem: greater demand is needed to drive economies of scale and reduce costs, but widespread adoption is hindered by current price premiums. It remains to be seen how the changing of economic, social and ethical incentives will alter these feedback loops. In particular scope-3 emissions reporting requirements, a rising carbon offset costs and more stable emissions trading markets are poised to tip the scales in favour of market growth in the realm of sustainable materials for the built environment.

Motivated by the prospect of future profitability through emissions reduction incentives, venture capital (VC) investments are increasingly driving the development and growth of sustainable materials ventures and their associated innovations. However, as we have seen above, venture capital flows are themselves subjected to silo-confinement, being obliged to deliver short-term returns in terms of portfolio growth and successful exits, thus neglecting integrative ventures that add community value. Moreover, while defined investment sectors of "Green-Tech", "ConTech", and similar labels have shown growing interest, investment in the sustainable construction sector is still relatively small compared to other areas of technology, with much of the venture capital entering this sector coming from venture arms of large corporate entities in the construction materials sector, such as Cemex Ventures, one of the leading investors in construction material startups. This can be attributed to barriers of higher capital expenditure and longer development times that sustainable construction ventures and other "hard-tech" enterprises face, when compared to those in sectors such as data analytics, software, or informatics.

As we have seen, due to feedback loops in market dynamics, sustainable material startups face difficulties scaling up production and competing with established materials producers. However, there are numerous focal points of growth and positive developments in the world of materials-oriented construction technologies. Motivated by integrative projects and networks that emphasise sustainability and systems thinking, such as those we have seen in the preceding section, materials innovators have been inspired to create various products and processes that produce more favourable sustainability outcomes, as indexed by schemes such as LEED, BREEAM, C2C and other sustainability certification schemes. There are several prominent trends in recent materials innovation aimed at improved sustainability metrics that are targeted or followed by startups and their VC backers. Some of these trends includes:

(1) **Biobased polymers**: the derivation of polymers such as polyurethane, polyhydroxyalkanoates (PHAs), polylactic acid (PLA) from waste biomass, or through the metabolism of micro-organisms, provides a pathway for the production of materials analogous or identical to those derived from fossil-hydrocarbons, with improved carbon-balance. By addressing particular niche applications of these materials, startups are able to gain market entry, through the provision of focused solutions in sustainability conscious construction projects. An illustrative example of this trend can be found in Klima-Pur windows, produced by Indresmat© in Spain. This is a bio-based polyurethane foam window frame with good recycling potential that provides high performance in thermal insulation, fostering energy-efficient buildings while providing a carbon-sink, as the carbon in the material come from the atmosphere carbon rather than fossil carbon [83,84].

(2) **Composites based on natural materials**: Novel engineered structural materials based on the adaptation of geomaterials or bio-based materials incorporated into various composite systems provide a pathway towards high performance with reduced environmental impacts. Examples include the use of cut stone laminated with carbon fibres, as demonstrated by Pierre-Carbone of Technocarbon and the use of high strength natural fibres (e.g. flax, jute or hemp) in composites for structural components. Hempcrete, a widely studied lime-based concrete reinforced with hemp fibres, and EWPs are also part of this trend, whereby the natural attributes of wood are harnessed in ways that are conducive to application-optimised mechanical performance. EWPs are also combined with other natural materials to create hybrid composites with improved sustainability and mechanical performance.

(3) **Robotic construction and 3D printing of materials**: The use of mechatronic approaches for production of structures is drawing increasing interest. A large part of this interest comes from the motivation of VC funds and innovators to apply technological solutions to housing shortages, rather than vernacular or community based solutions. The attractiveness of mechatronic approaches stems from economic factors, namely the possible reduction in labour costs and delivery times, as well as efficiency, as material wastage is reduced. Robotic fabrication and 3D printing enable precision and speed, facilitating the creation of intricate geometries that would be difficult to achieve with traditional methods. Ventures in this field seek to integrate innovations into the equipment, formulations and incorporation of waste feedstocks.

(4) **Advanced nanomaterials**: Despite the seemingly stark contrast between the costs of nanoparticles and conventional structural materials, the integration of advanced nanoparticles in concrete has been demonstrated as a pathway towards synergetic outcomes. Of particular note is the use of graphene, or more accurately graphene oxide, as an additive in concrete [85-89]. By catalysing hydration, nucleating high density calcium silicate hydrates and by bridging cracks, small levels of these additives can produce dramatic improvements in the performance of the cementing phase in concrete, enabling the use of lower amounts of cement and providing a pathway to lower embodied emissions in the material. In general, the integration of advancements in nanotechnology in materials for the built environment is an area of high growth with numerous ventures active in this sector.

(6) **Biotechnology**: Around the globe, biotechnology ventures have seen rapid growth in recent years. Towards sustainable materials for the built environment, the integration of disciplines of molecular biology and genetic engineering has allowed the harnessing of various microscopic organisms including yeast, plankton, algae and fungi towards the production of improved materials based on mycelium, bio-based polymers, and biocementation of minerals.

(7) **Green concrete**: The realisation of opportunities for industrial symbiosis underpins many emerging enterprises in the field of sustainable construction materials. Bringing the right waste streams and processes into place to facilitate high performance concrete with better sustainability outcomes has enabled a plethora of ventures broadly categorised as Green Concrete. This includes cements and SCMs based on slags, fly ash and other industrial by-products.

The representative trends described above are by no means an exhaustive list of the numerous fields of innovation supported by capital investments. It can be seen that these trends are all in some way examples of cross-disciplinary integration, bridging fields of biotechnology, mineralogy, agriculture and many more. Despite the numerous fields of innovation, there are some significant limitations to the ability of VC funds to promote sustainable built environments. VCs often seek rapid returns on investment, which can be challenging in the construction industry where project cycles are long and adoption of new materials can be slow. Moreover, the profitability and high expected internal rates of return (IRRs) that are required for recipient ventures excludes enterprises seeking to develop holistic solutions that do not have pathways to high revenues. Thus for example earth and stone construction, vernacular building techniques, regenerative agriculture and other approaches that are not conducive to patenting, or establishment of exclusive supply chains may not gain momentum under such funding models.

Venture capital and corporate venture capital (CVC) clearly play a critical role in funding innovative and sustainable approaches in construction materials, enabling agile responses to emissions regulations and growing demand for eco-friendly alternatives. However, the profit-driven nature of these ecosystems may hinder cross-disciplinary integration. Venture capital prioritises investments that promise high returns and market exclusivity, which can limit the scope of integration of disciplines. As a result, cross-disciplinary integration is typically fostered by investors only when it offers a unique, proprietary advantage that ensures competitive differentiation. This focus on exclusivity and economic efficiency can inadvertently stifle broader, more inclusive innovation, as projects that require open collaboration or lack immediate commercial potential may struggle to secure funding. Consequently, while VC and CVC ecosystems drive progress in sustainable construction materials, their inherent constraints can hinder the full potential of cross-disciplinary solutions needed to address complex global challenges.

Economic and financial constraints to cross disciplinary integration are further compounded by regulatory frameworks that may lack the flexibility to support approaches that span multiple disciplines or industries.

*Regulatory Frameworks and Their Limitations*

Regulatory frameworks in the built environment are traditionally developed within specific disciplinary boundaries, focusing on core objectives such as safety, structural integrity, and basic environmental compliance. While these regulations are needed to ensure baseline performance, they often lag behind the pace of innovation and fail to incentivise cross-disciplinary collaboration, which is critical for advancing sustainable material use. For example, while building codes frequently mandate minimum energy efficiency standards, these may not sufficiently reward materials or processes that exceed these standards [90,91].

Regulatory environments tend to be rigid, lacking adaptability and often prioritise compliance over innovation or excellence in sustainability. This is rarely conducive to the exploration of disciplinary integration towards outstanding sustainability outcomes. Building codes and standards are generally slow to incorporate advancements in sustainable materials, such as bio-based composites, mycelium derived materials or other innovations in construction materials. This can create uncertainty for architects, engineers, and developers who may be hesitant to adopt these materials due to concerns about compliance, delays or liability [92,93]. Additionally, the absence of unified global protocols for measuring and certifying environmental performance—such as embodied carbon or circularity potential—further complicates efforts to align regulatory frameworks with sustainability goals. Without clear guidelines, professionals are often left to navigate a patchwork of local, national, and international regulations, which can stifle innovation and collaboration across silo and national borders.

The regulatory framework may pose particular obstacles with regard to the use of natural materials, which often exhibit variable material properties. In particular, the use of earthen construction techniques—such as rammed earth, adobe, and cob—can face significant hurdles due to building code compliance requirements[94-96]. These materials, while offering pathways to high performance in terms of hygroscopic temperature regulation, thermal mass, and low embodied carbon [97-102], may face challenges meeting the standardised performance metrics required by building codes and guidelines. This disconnect between the inherent benefits of natural materials and rigid regulatory standards can stifle innovation and limit their adoption, despite their potential to contribute to sustainable, resilient, and energy-efficient built environments. Addressing these challenges will require updated regulatory frameworks that account for the unique properties and performance of natural materials, fostering their integration into mainstream construction practices.

Gradually, regulatory frameworks are beginning to encourage cross-disciplinary integration. For instance, the European Union's Circular Economy Action Plan from 2020 includes measures to promote the use of recycled materials in construction and mandates building deconstruction over demolition, fostering collaboration between material scientists, designers, and contractors [103].

**Tools for disciplinary integration**

Many modelling and analytical tools developed for design in the built environment embody some measure of cross-disciplinary integration; Finite element analysis (FEA) can simulate the structural and thermal performance in materials, integrating multi-physics understandings in the context of architecture and engineering [104]. Importantly, for the analysis of lifecycle impacts of applied materials in the built environment, LCA tools including software like GaBi and OpenLCA help bring together inputs from the silos of diverse stakeholders and combine them to produce actionable insights into the environmental impacts and sustainability outcomes [105,106]. A new emerging tool to enhance the integration of industries and disciplines are dynamic LCAs, and real-time LCAs, that take into account the variation of lifecycle inventories over time, and can be updated to reflect changes across various nodes in the supply chain.

Emerging digital tools, including neural networks and artificial intelligence (AI), are revolutionizing cross-disciplinary integration in the built environment by enabling seamless collaboration and data-driven decision-making. AI-powered platforms, such as generative design tools, allow architects, engineers, and material scientists to explore countless design iterations optimised for sustainability, performance, and cost. These tools leverage machine learning to analyse complex datasets, identifying patterns and solutions that might be overlooked in traditional workflows [107,108] . For example, AI can optimise material selection by balancing factors like embodied carbon, structural performance, and availability, fostering collaboration between disciplines. Moreover AI tools are fostering cross-disciplinary collaboration by enabling the rapid exploration of many dimensions of engineering challenges, empowering architects, engineers, and material scientists to holistically evaluate sustainable material design, integration, and performance across the entire lifecycle of built environments.

A particularly important tool facilitating the integration of disciplines is Building Information Modeling (BIM), which provides a shareable digital representation, often in a format compatible across multiple platforms that enables stakeholders to collaborate in real time to improve outcomes regarding the construction and use of structures. BIMs bring together the various elements of a building and its materials in a standardised format that can be used to optimise design and use of materials and structures throughout the lifetime of the structure. Construction, Energy use, maintenance, utilities, renovation and other aspects of a building can be included in the information of BIMs. In recent years, BIMs are becoming a mandatory requirement for projects over a certain size [109]. Integrating BIMs with artificial intelligence and Internet of things (IoT) offers pathways to create integrative outcomes in construction and urban development [110,111].

**Cross-disciplinary integration and systems thinking in the context of Vernacular Construction**

In the context of vernacular construction, cross-disciplinary integration is playing a growing role. In earth based construction, disciplines of granular physics, mineralogy, agriculture, architecture, materials science and microbiology are brought together through the manual labour of craftsmen to form sustainable vernacular structures [94]. The use of local materials in construction, including earth, stone, wood and waste biomass, requires the integration of insights from local experts as well as discipline professionals. Improving and expanding the field of vernacular construction may benefit from further platforms and channels for discourse and collaboration.

3D printing and digitisation have been applied to earthen construction, posing an inspiring example of cross-disciplinary integration [112-115]. By combining advanced fabrication technologies with traditional earth-based techniques, researchers and practitioners are creating innovative solutions that enhance structural performance, design flexibility, and scalability. For instance, 3D-printed earthen structures leverage computational design to optimise material use and structural integrity, while maintaining the sustainability and cultural relevance of vernacular methods. This fusion of ancient wisdom and modern technology not only preserves the ecological benefits of natural materials but also opens new avenues for affordable, resilient, and aesthetically rich construction. Such advancements underscore the transformative potential of cross-disciplinary collaboration in redefining sustainable building practices for the future.

**Conclusions and outlook**

As we have seen, the integration of diverse disciplines is essential for advancing sustainable construction practices addressing the complex challenges of resource efficiency, environmental impact, and human well-being. Sustainable construction requires the collaboration of architects, engineers, material scientists, environmentalists, educators, and policymakers to ensure that every stage of the building lifecycle—from material extraction and production to construction, operation, and end-of-life disposal—is optimised for sustainability. By breaking down the silos that traditionally separate these fields, cross-disciplinary integration can foster innovative solutions that balance technical performance, environmental stewardship, and social equity. The incorporation of LCA tools into design processes allows architects and engineers to evaluate the environmental impacts of material choices, while input from environmental scientists can guide the selection of low-carbon alternatives. This collaborative approach may reduce the carbon footprint of buildings and further enhance their resilience and adaptability to changing environmental conditions.

Towards sustainable outcomes the integration of disciplines must extend beyond technical fields to include social sciences, economics, and community engagement, ensuring that construction practices are culturally relevant and socially inclusive. For example, vernacular construction techniques, which have historically relied on local materials and traditional knowledge, can be revitalised through the integration of modern technologies such as 3D printing and digital fabrication. This fusion of ancient wisdom and contemporary innovation not only preserves cultural heritage but also promotes the use of sustainable, locally sourced materials. By fostering systems thinking and shared visions among diverse stakeholders, discipline integration can drive the transition towards a more sustainable and equitable built environment, where innovation and tradition coexist to meet the needs of present and future generations.